\documentclass[a4paper,11pt]{article}
\usepackage{pos}

\title{Performance and extensions of the central data acquisition system for Phase$\,$II of the Pierre Auger Observatory}
\ShortTitle{Cosmic Ray Indirect - \#140}

\author*[ab]{Paul Filip}

\affiliation[a]{Institute for Astroparticle Physics, Karlsruhe Institute for Technology\\
  Hermann-von-Helmholtz Platz 1, 76344 Eggenstein-Leopoldshafen, Germany}
\affiliation[b]{Instituto de Tecnologias en Deteccion y Astroparticulas, Universidad Nacional de General San Martín\\
Av. Gral. Paz 1499, B1650 Buenos Aires, Argentina}

\onbehalf{for the Pierre Auger Collaboration$^c$}
\affiliation[c]{Observatorio Pierre Auger, Av.\ San Mart{\'\i}n Norte 304, 5613 Malarg\"ue, Argentina\\
{Full Author List \rm\url{https://www.auger.org/archive/authors_icrc_2025.html}}}

\emailAdd{spokespersons@auger.org}

\abstract{The Pierre Auger Observatory is a hybrid detector designed to observe and study ultra-high-energy particles of extraterrestrial origin. With its 27 fluorescence telescopes and over 1600 autonomously operating water-Cherenkov detectors spread over an area of \SI{3000}{\kilo\meter\squared}, it is world-leading in terms of exposure to cosmic rays and offers an unparalleled window into the physical processes that happen at energy scales unattainable by particle accelerators on Earth.

Measurement information of candidate air-shower events from all associated detectors and telescopes is collected at a central data-acquisition system located in the nearby town of Malargüe, and processed for higher-level physics analysis. On top of this, data for monitoring the long-term stability and operation of the observatory is forwarded to the central server as well.

In this work, we briefly review the central data-acquisition system of the Pierre Auger Observatory. We examine the rates, efficiencies, and purity of detected events for Phase$\,$II of the Pierre Auger Observatory and compare them to performance parameters during Phase$\,$I. We detail challenges in the event detection up until now and present recent changes in the central data acquisition system and local station software that aim to streamline the data acquisition chain.}

\ConferenceLogo{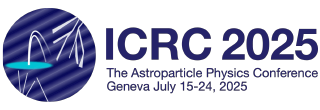}

\FullConference{39th International Cosmic Ray Conference (ICRC2025)\\
 15–24 July 2025\\
Geneva, Switzerland\\}

\begin{document}
\maketitle

\begin{figure}[h!]
  \centering
  \def\h{0.36}
  \subfloat[]{\includegraphics[height=0.35\textwidth]{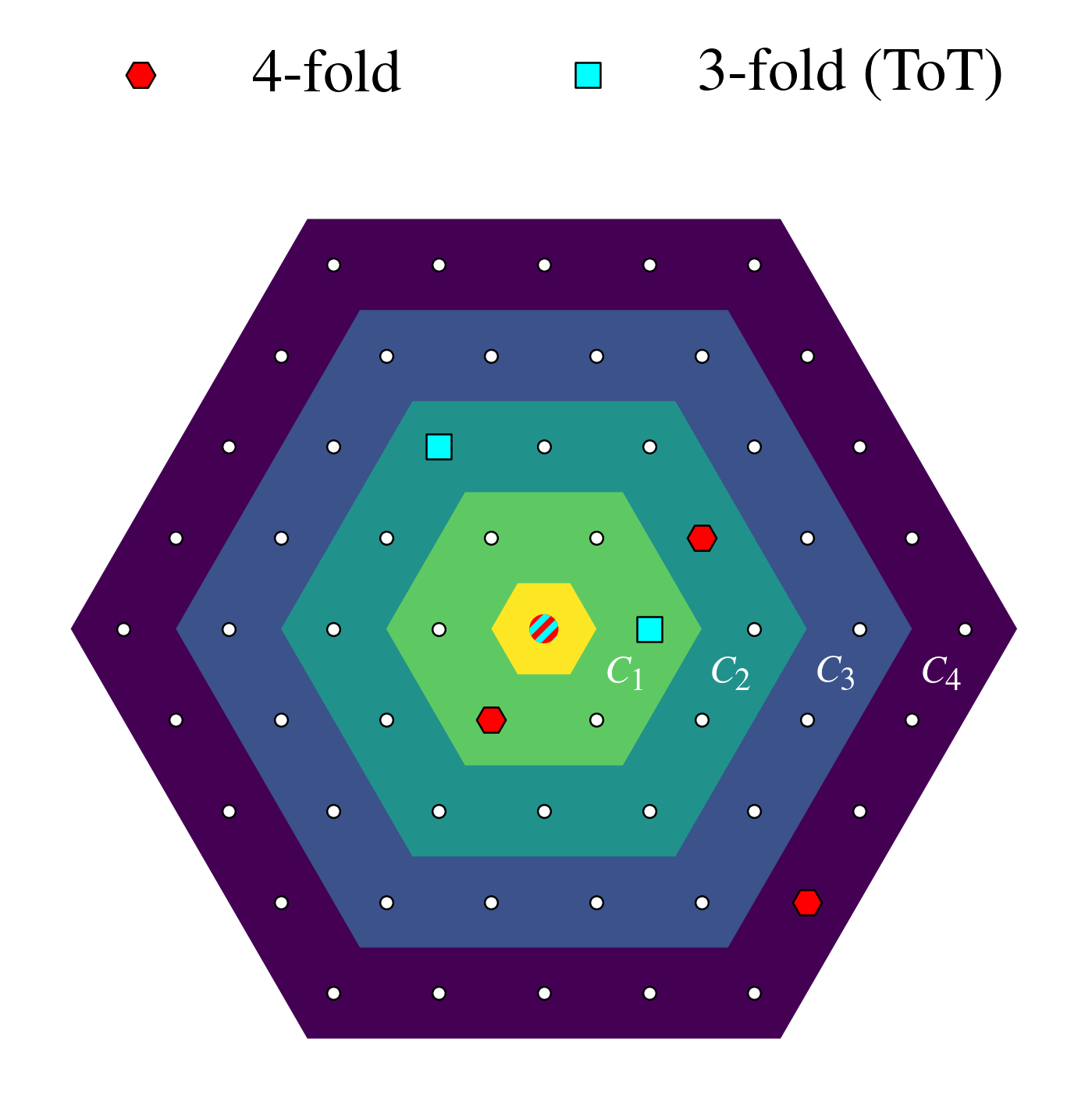}
  \label{fig:t3-modes}
  }\hfill
  \subfloat[]{\includegraphics[height=0.35\textwidth]{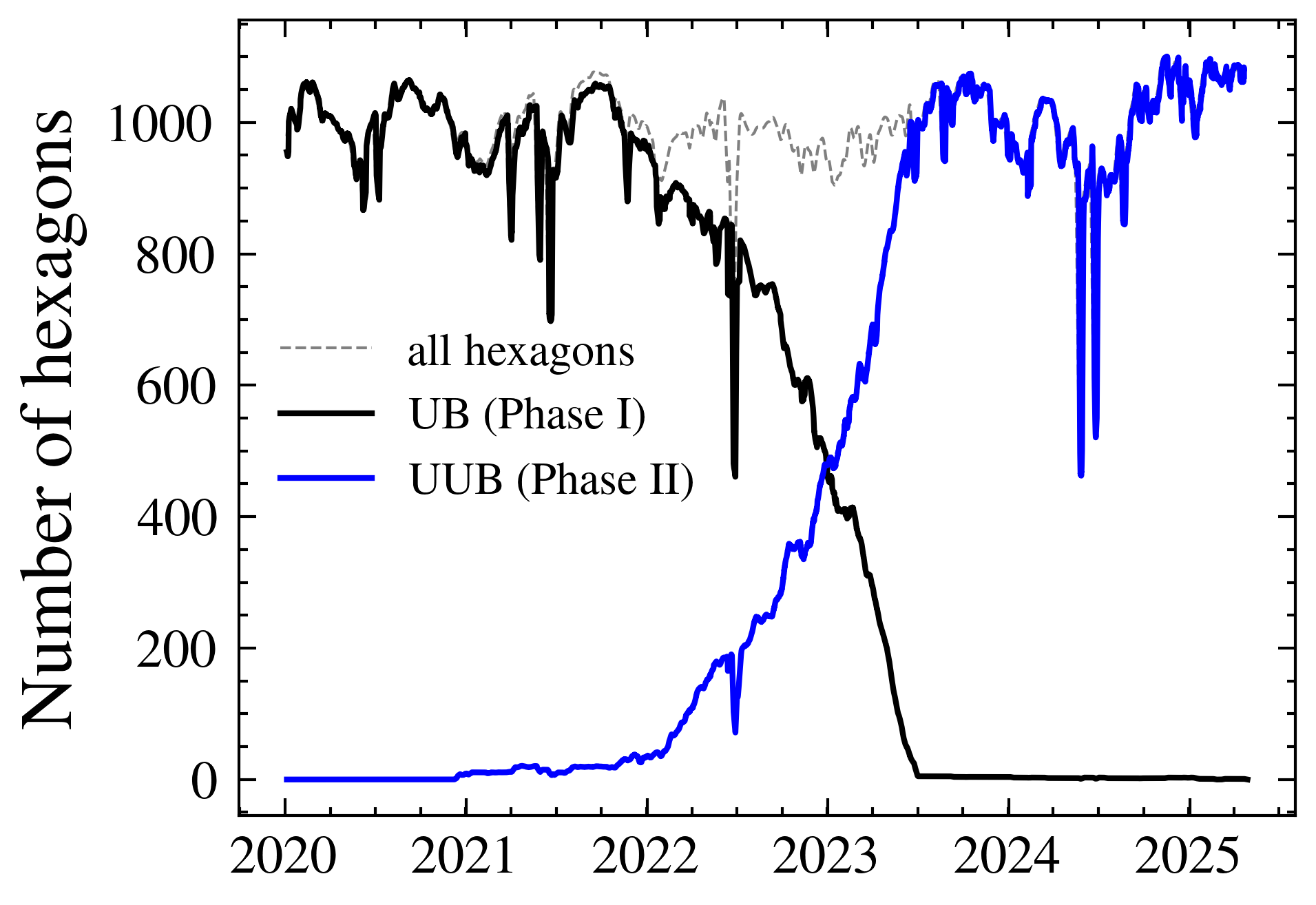}
  \label{fig:hexagons}
  }
  \caption{\subref{fig:t3-modes} The event trigger is based on the $n$ nearest neighbors, organized in hexagonal rings around a central triggered station. For the 3-fold trigger $2\mathrm{ToT}C_1\&3C_2$ (cyan), at least one nearest neighbor (in $C_1$), as well as one next to nearest neighbor (in $C_2$), must be present. On top of this, only T2 triggers of a specific type are considered during pattern matching. The $2C_1\&3C_2\&4C_4$ (red) considers all station-level triggers but requires at least one additional station that can be as far as a 4th nearest neighbor. \subref{fig:hexagons} Weekly average number of hexagons during the AugerPrime upgrade, split into upgraded stations (blue), and stations with original electronics (black). Bad weather during the winter months is responsible for large-scale array shutdowns.}
\end{figure}

\section{Introduction}

The Pierre Auger Observatory covers an area of roughly \SI{3000}{\square\kilo\meter} and is the world's largest detector for extensive air showers stemming from cosmic rays of ultra-high energies. The observatory consists of 27 UV-sensitive telescopes, called the Fluorescence Detector (FD), and $>1600$ stations that comprise the Surface Detector (SD). The immense size of the observatory, coupled with its remote location in the Argentinian pampa, poses unique challenges. For example, a single SD station collects data with a sampling rate of \SI{120}{\mega\hertz}, which results in a theoretical bandwidth of the order of \unit[per-mode=symbol]{\tera\byte\per\second} for the SD alone. Of course, it is not feasible to transfer this amount of data over the distances present in the SD array. Instead, data from a station must pass through a hierarchical trigger system before being sent to a Central Data Acquisition System (CDAS). In this way, the data transfer rate\footnote{Including also diagnostic information about the detector performance.} between the CDAS $\leftrightarrow$ station can be limited to $<\SI[per-mode=symbol]{150}{\byte\per\second}$. An exact discussion of the communication system and the software implementation is given in Ref. \cite{satoProceeding}.

The lowest level in the trigger hierarchy is the station level. Every SD station independently monitors its detector data for several trigger conditions. If a trigger condition is met, the station sends a station-level trigger (T2) with a microsecond timestamp to the CDAS. The CDAS scans all incoming T2s for spatio-temporal coincidences, which are usually caused by air showers induced by cosmic rays. Two patterns are considered when searching for coincidences: the 3-fold $2\mathrm{ToT}C_1\&3C_2$ pattern and the more permissive 4-fold $2C_1\&3C_2\&4C_4$ pattern (ref. \Cref{fig:t3-modes} and \cite{sdtriggerpaper}). If such a pattern is recognized, the CDAS issues an event-level trigger (T3) and requests data from the participating triggered stations to collect all information related to an event.

With the AugerPrime upgrade, the Pierre Auger Observatory achieves a higher discrimination on the primary particle mass and enters Phase$\,$II of its data collection. During the upgrade, all of the SD station electronics have been upgraded from a Unified Board (UB, 10-bit ADC sampled at \SI{40}{\mega\hertz}) to the Upgraded Unified Board (UUB, 11-bit ADC sampled at \SI{120}{\mega\hertz}). 

With the higher resolution of the UUB data, the event detection chain changes in principle from the ground up. However, steps have been taken to ensure the backward compatibility of Phase$\,$I and Phase$\,$II triggers \cite{augerprime}. We show this in \Cref{sec:performance} by compiling a preliminary CDAS performance report for Phase$\,$II and comparing it with the performance during Phase$\,$I. We highlight challenges present in the event detection chain and discuss how they can be solved from \Cref{sec:challenges} onward. 

\section{CDAS Performance Phase$\,$I + Phase$\,$II}
\label{sec:performance}

\begin{figure}
  \centering
  \subfloat[]{\includegraphics[height=0.24\textwidth]{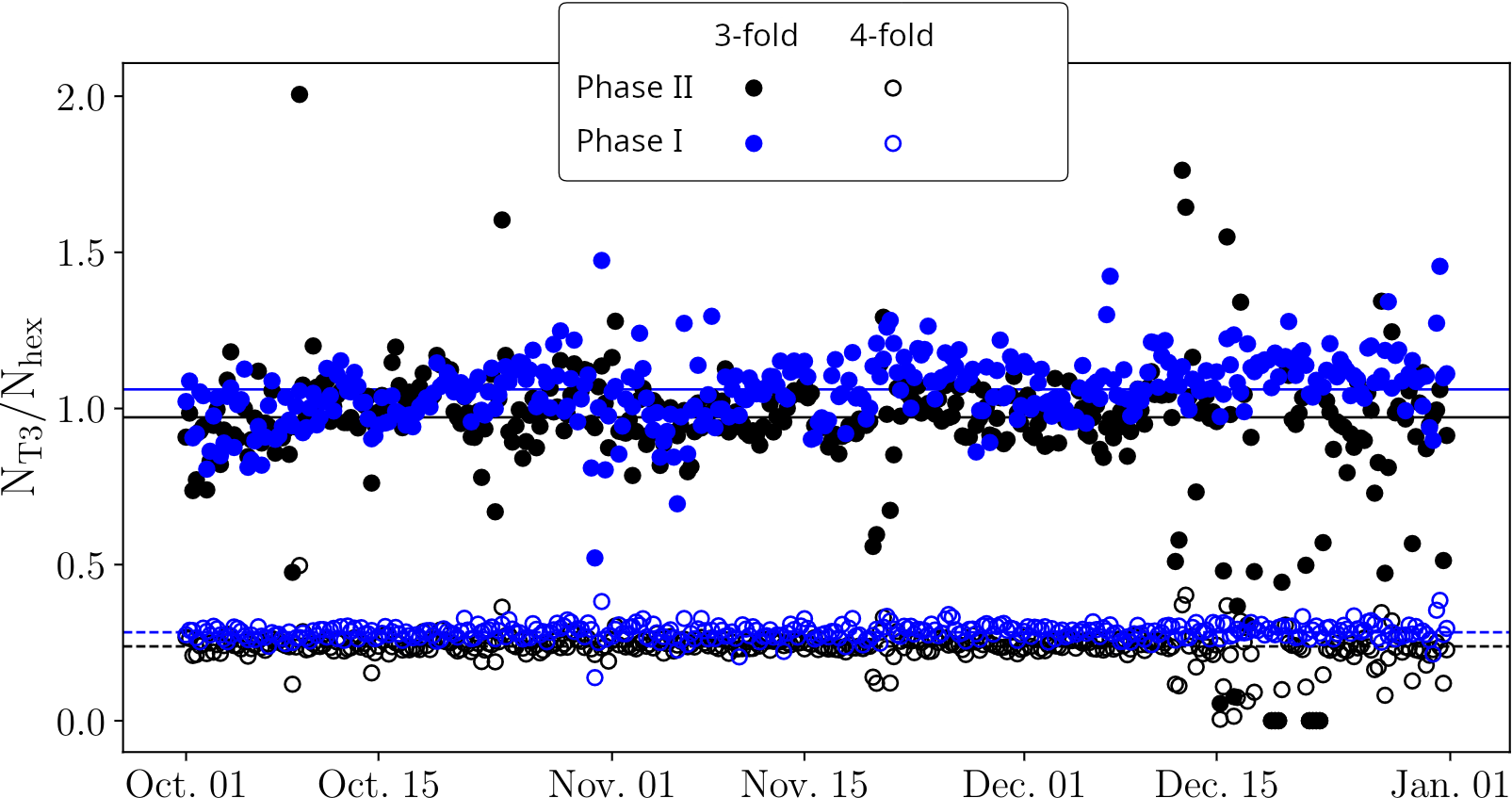}
  \label{fig:t3-rate}
  }\hfill
  \subfloat[]{\includegraphics[height=0.24\textwidth]{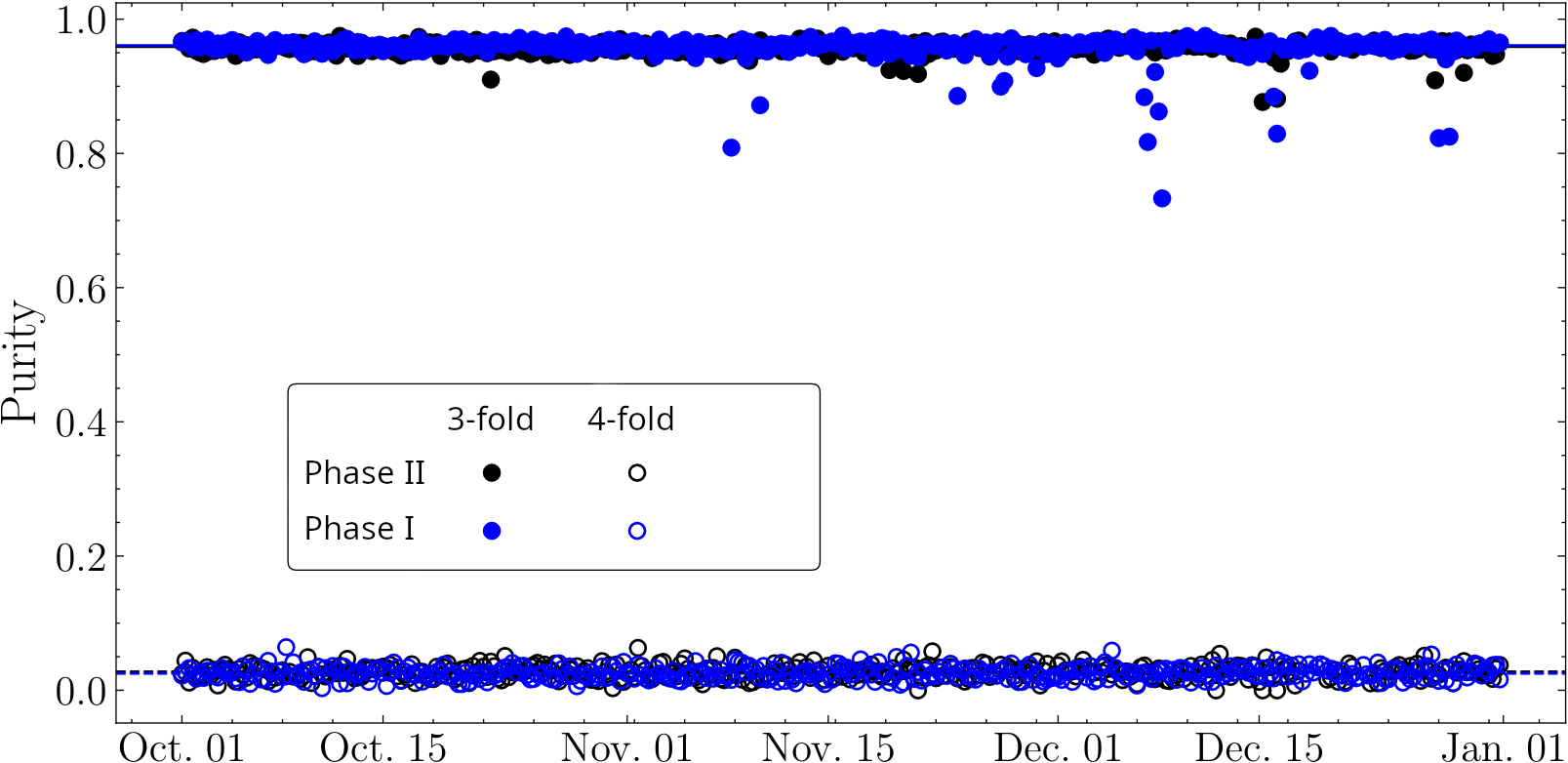}
  \label{fig:t3-purity}
  }
  \caption{\subref{fig:t3-rate} Number of T3s per hexagon, separated by trigger mode for Phase$\,$I data (blue) and for Phase$\,$II data (black). The Phase$\,$I data was collected from October 2021 to December 2021. The Phase$\,$II data stems from the same timespan in 2024 and displays high variance at the end of the DAQ period, likely due to thunderstorms. \subref{fig:t3-purity} The T3 purity of the aforementioned dataset. Horizontal lines give the mean T3 purity.}
\end{figure}

\begin{table}[h!]
\begin{minipage}[t]{0.48\linewidth}

A number of parameters have proven useful to monitor the low-level performance of the CDAS. For example, the T3 frequency should resemble the rate with which air showers impinge on the SD array. Since the latter is constant for UHECRs, the former should be (with invariant trigger settings) too constant. Of course, this is assuming a detector of constant area, which was not the case for the SD during Phase$\,$II commissioning (cf.\ \Cref{fig:hexagons}). 

\end{minipage}\hfill
\begin{minipage}[t]{0.48\linewidth}
\vspace{-0.3cm}
\caption{CDAS performance parameters. Event rates are given in no. of T3s per hexagon per day.}
\label{tab:lorenzo-analysis}
\begin{tabularx}{\textwidth}{lccr}
\toprule
& Phase$\,$I & II (prelim.) & $\Delta$ / \%\\
\midrule
3-fold rate & $4.24$ & $3.89$ & $-8$ \\
3-fold purity & $0.961$ & $0.960$ & $\pm0$ \\
\midrule
4-fold rate & $1.13$ & $0.95$ & $-16$ \\
4-fold purity & 0.025 & 0.028 & +9 \\
\midrule
SD uptime & $0.96$ \cite{sdtriggerpaper} & $0.957$ & $\pm0$ \\
\bottomrule
\end{tabularx}

\end{minipage}%
\end{table}

The number of SD unit cells, called hexagons, consisting of one central station and its six nearest neighbors increases (decreases) for the UUB (UB) array as individual stations are upgraded. As the UUB array grows, it detects more air shower events. We therefore define \emph{T3 rate} as the frequency of (types of) T3s, normalized to the number of hexagons that operate simultaneously in the SD array. 

After registering a T3 and requesting event data from stations, the CDAS also applies physics-based quality checks, storing their outcome alongside the event information. These checks filter out random coincidences of T2s and deliver T3 events that are almost entirely formed by extensive air showers. Thus, a measure of the background contamination in the T3 dataset is given by the fraction of events that pass the higher-level selection criteria. This is the \emph{T3 purity}. Lastly, we quantify the duty cycle of the SD array during an observation period by splitting it into intervals of \SI{15}{\minute}. We build the \emph{SD uptime} by dividing the number of intervals that do not have T3 events by the number of all intervals.

To compare the CDAS performance between Phase$\,$I and Phase$\,$II, we collect all T3 events from October to December in the years 2021 and 2024. We choose the specific periods due to the good weather\footnote{Thunderstorms and heavy rain were reported over the observatory at the end of 2024.} and the overall stable operation of the SD. Moreover, the UB array in 2021 and the UUB array in 2024 have a comparable size and thus exposure (cf.\ \Cref{fig:hexagons}). 

This serves as a tentative analysis to gauge the CDAS performance in Phase$\,$II. Quality cuts have not been applied to the data. The systematics of the analysis have not been estimated but are assumed to be of the order of $\SI{10}{\percent}$. In any case, the results presented in this section are meant to be understood as preliminary. The calculated T3 rate and purity for Phase$\,$I and Phase$\,$II are displayed in \Cref{fig:t3-rate} and \Cref{fig:t3-purity}. Their mean values and the SD uptime over the entire DAQ period are also listed in \Cref{tab:lorenzo-analysis}. 

We observe a slight decrease in T3 rates for both the 3-fold and the 4-fold trigger modes. The difference in the performance of Phase$\,$I and Phase$\,$II is probably due to the compatibility mode in which the SD stations operate in Phase$\,$II. In compatibility mode, the UUB emulates the UB electronics readout process and applies the same station-level trigger algorithms to the filtered and downsampled data that were in use in Phase$\,$I. Evidently, the emulation is not perfect, as the T3 rates are not backwards compatible and overall, fewer events are observed. 

The T3 purity remains unchanged for the 3-fold trigger and increases slightly for the 4-fold trigger. The SD uptime is very similar for both DAQ periods and implies that no significant downtimes were observed during the detector upgrade, both for the UB and UUB array. 

\section{Sources of transient noise}
\label{sec:challenges}

\begin{figure}
  \centering
  \subfloat[]{\includegraphics[height=0.32\textwidth]{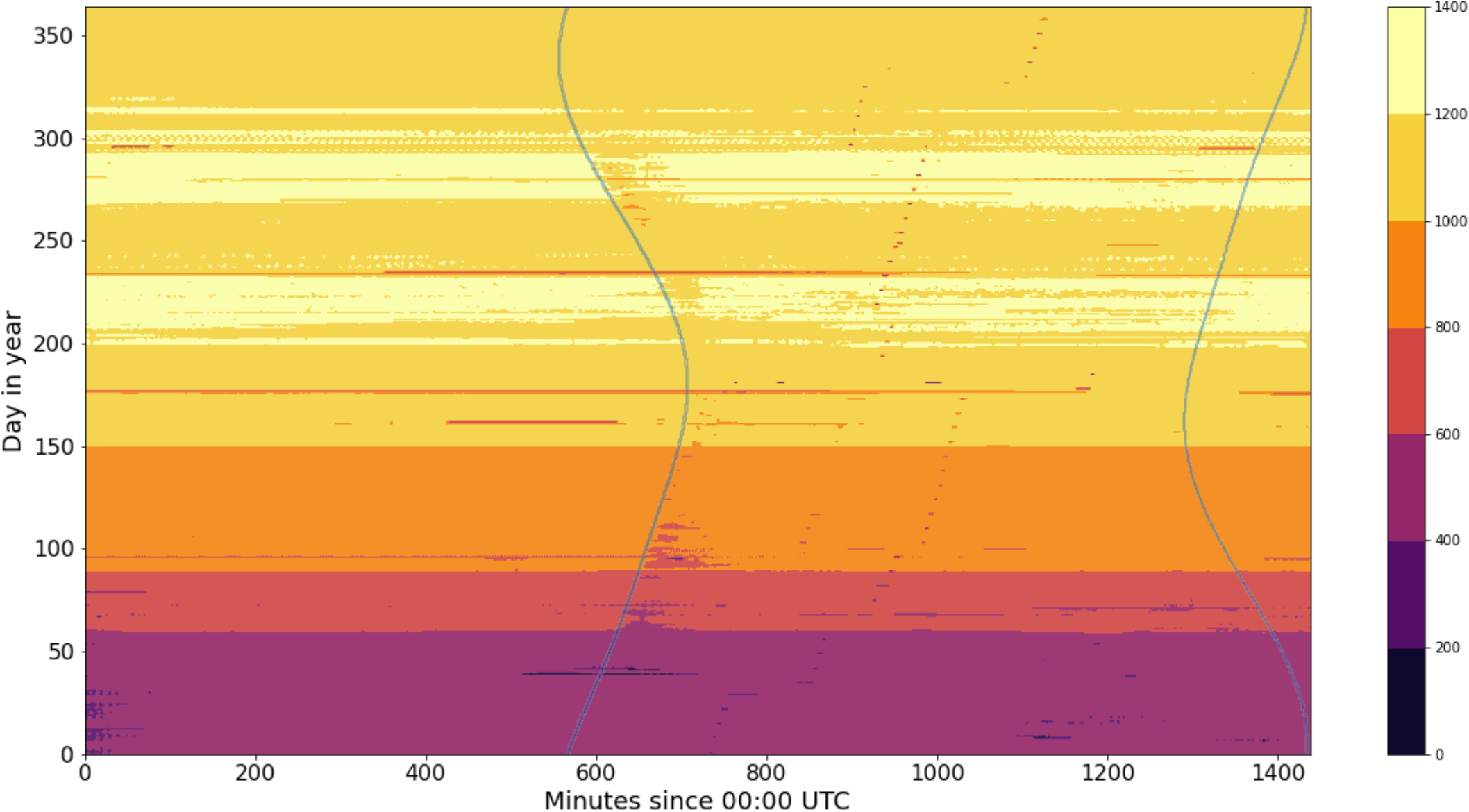}
  \label{fig:sunrise-noise}
  }\hfill
  \subfloat[]{\includegraphics[height=0.32\textwidth]{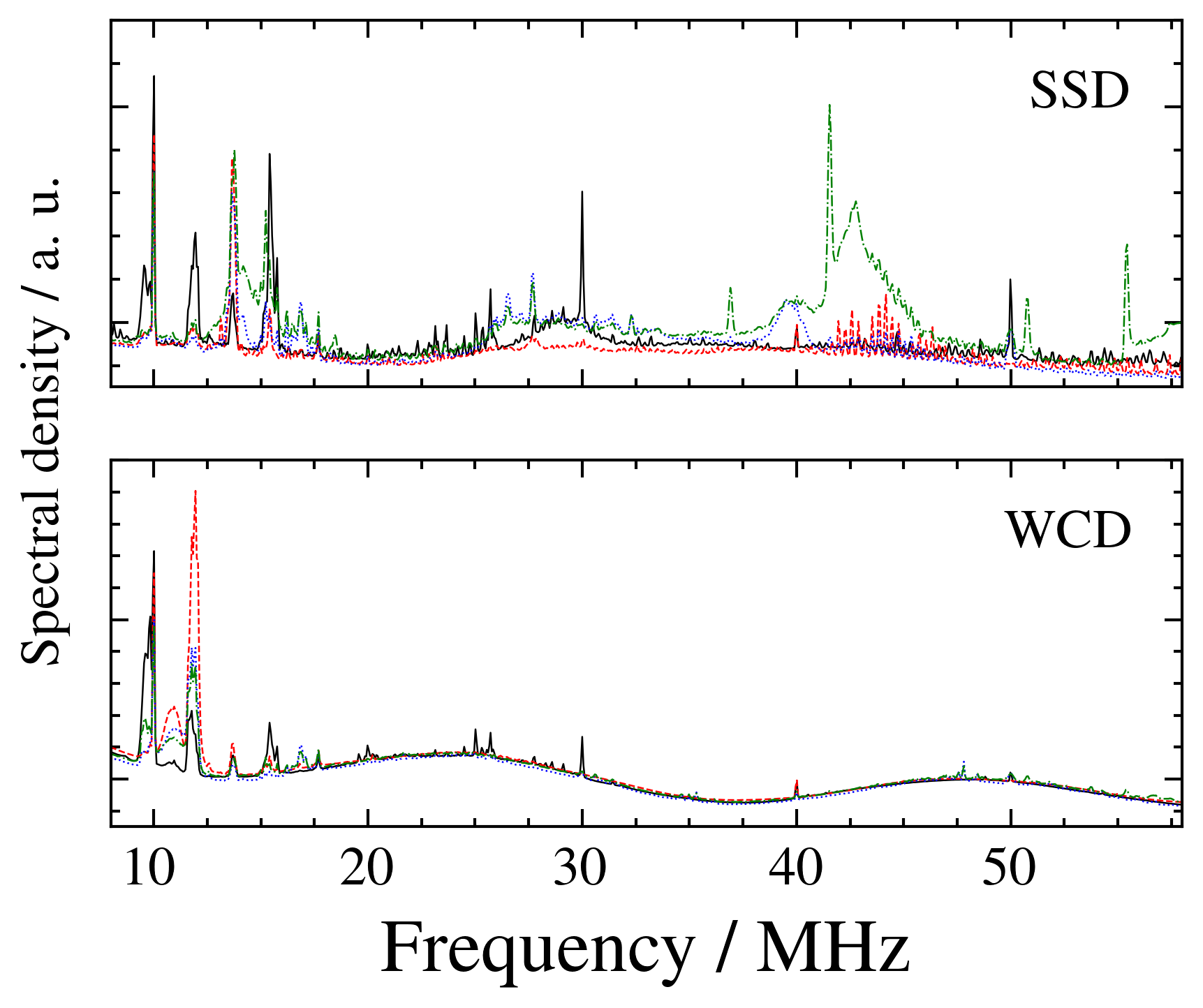}
  \label{fig:power-spectrum}
  }
  \caption{\subref{fig:sunrise-noise} Number of fully functional hexagons in the UUB array for 2023. The sinusoidal blue lines mark the time of sunrise and sunset. \subref{fig:power-spectrum} The spectral density of the SSD (top) and WCD (bottom) detector for background data of various UUB stations. Several sharp peaks demonstrate the systematic influence of electronic noise. The data was recorded in 2022-11 (black, red) and 2023-03 (green, blue).}
\end{figure}

Several sources of noise are and have been present in the T2/T3 datasets for Phase$\,$II. Although the underlying effects and problems have, in part, been known since Phase$\,$I already, their impact is amplified with the advent of the higher data resolution that the new station electronics offer.

\subsection{Local Sidereal Time}

\Cref{fig:sunrise-noise} shows the number of working UUB hexagons in the array for the year 2023. A very careful inspection of the data reveals a pattern that persists throughout the year. Each day, shortly after sunrise, the number of functioning hexagons drops by up to $40\%$, causing a decrease in exposure of 0.1\% over the entire year on average. This is caused by individual stations reporting a very high T2 rate, which results in the CDAS issuing a reset command in the station and temporarily removing it from the active DAQ (Ref. \cite{carlamonitoring}). 

As such, the noise present in the stations is only indirectly observed via a degraded performance of the SD. The effect is clearly related to the local sidereal time in the UUB array. The Tank Power Control Board (TPCB) has been theorized to be responsible for the burst in the trigger rate when charging station batteries during the morning hours. The problem can be mitigated in the laboratory by equipping the TPCB with diodes and capacitors. With all TPCBs in the field being upgraded in this fashion and the implementation of the trace-cleaning algorithm in \Cref{sec:cleaning}, the behavior is no longer observed.

\subsection{Electronic Noise}

To gauge the presence of electronic noise in the detectors, we analyze background data of the SD. The data set on hand originates from two different measurements in November 2022 and March 2023 and contains \SI{394.4}{\second} of randomly sampled detector data collected from four stations. It is divided into sets of 2048 consecutive samples that measure $\SI{17}{\micro\second}$ in duration. During the DAQ, 5000 such traces are acquired in quick succession and written to disk. The writing introduces a considerable delay, resulting in roughly \SI{60}{\second} of data being stored in a \SI{22}{\hour} interval.

We calculate the FFT of the data on a trace level, applying a Hanning window to minimize spectral leakage. \Cref{fig:power-spectrum} shows the mean FFT, grouped by SSD and WCD, for all four stations.

\begin{figure}
    \centering
    \includegraphics[width=1\linewidth]{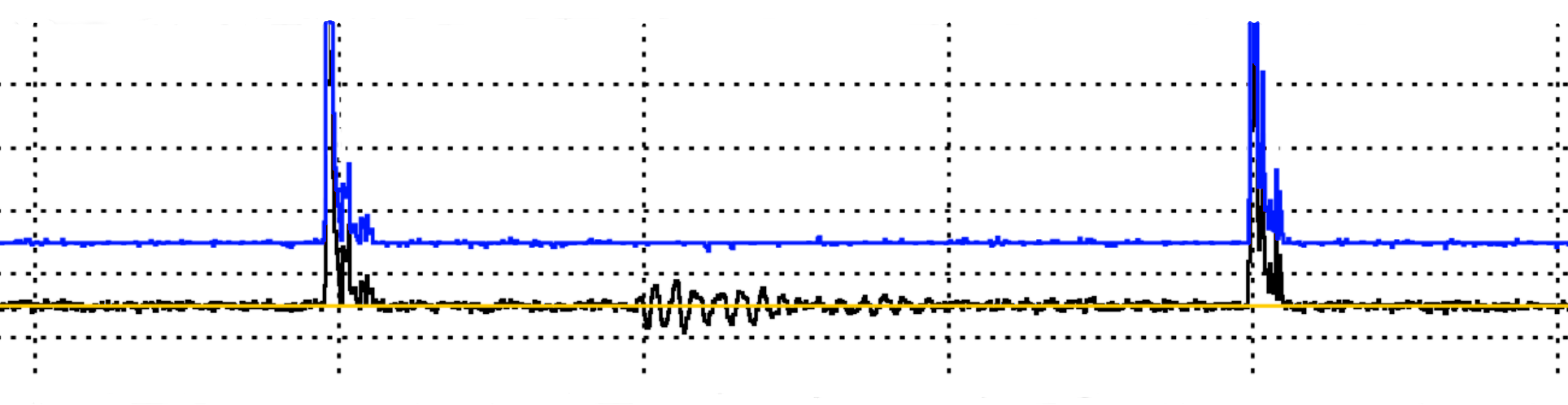}
    \caption{An erroneously triggered WCD trace (black) displays electronic noise in the form of a $\SI{10}{\mega\hertz}$ oscillation around the baseline (orange) between two sub-threshold muon pulses. The blue trace is the output of a trace-cleaning algorithm using the black trace as input. The effect of the electronic noise is mitigated.}
    \label{fig:trace-cleaning}
\end{figure}

Several frequency bands are clearly polluted by artificial signal constituents. In unfortunate circumstances, electronic noise can trigger a T2 trigger, as shown in \Cref{fig:trace-cleaning}. The erroneously sent T2 can then form a T3 pattern with coincident T2s at neighboring stations. This results in the false detection of an air shower event and decreases the T3 purity of the CDAS.

\subsection{Thunderstorms}

\begin{figure}
  \centering
  \subfloat[]{\includegraphics[height=0.33\textwidth]{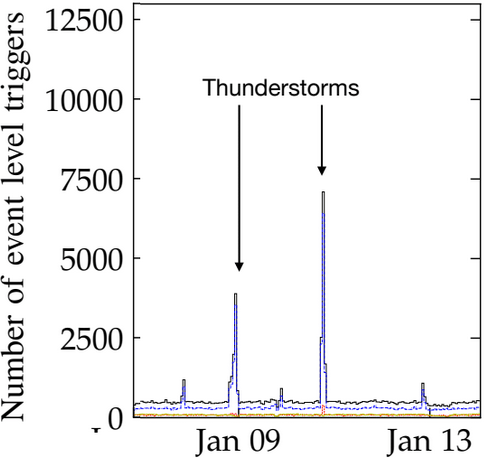}
  \label{fig:t3-rate-thunder}
  }\hfill
  \subfloat[]{\includegraphics[height=0.33\textwidth]{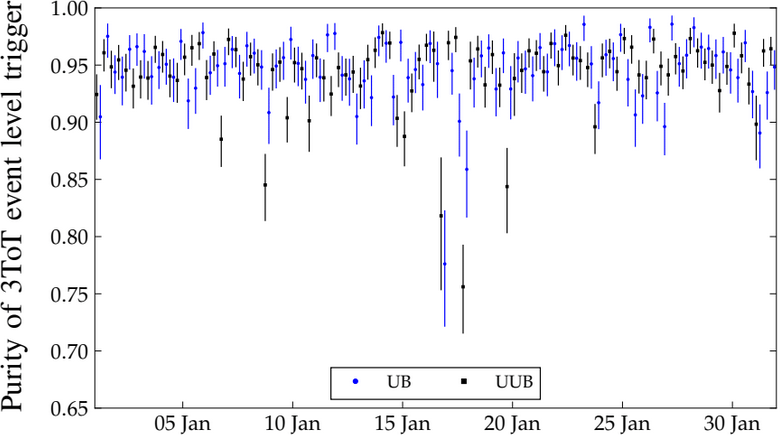}
  \label{fig:t3-purity-thunder}
  }
  \caption{\subref{fig:t3-rate-thunder} A large number of 3-fold T3s is recorded during periods with thunderstorms, as indicated by the arrows. \subref{fig:t3-purity-thunder} The corresponding 3-fold T3 purity rate measures a drop coincident with the weather period.}
  \label{fig:thunderstorm}
\end{figure}

The largest source of detector instabilities is lightning activity. Both in Phase$\,$I and Phase$\,$II, a spike in the T3 rate and a subsequent drop in the T3 purity are observed during and in the wake of thunderstorms, as shown in \Cref{fig:thunderstorm}. During these periods, the SD stations exhibit an elevated rate of T2 triggers. More importantly, T2 triggers from neighboring stations are highly correlated in time, as they are likely caused by the same electric discharge. Consequently, the CDAS identifies many more T3 events than during nominal operation. Since the communication between the CDAS and SD stations is bandwidth-limited, only one to two T3 requests can be issued to the SD array every second. This represents a bottleneck that results in the buildup of a large T3 queue during thunderstorms. The CDAS naively iterates through the T3 queue and requests data from a station often hours after the event happened.\footnote{The SD stations save T2 data for $\sim\SI{30}{\second}$.} Such stale events block the acquisition of new, real air shower events and drop the T3 purity and SD uptime considerably. The estimated effect on the duty cycle of the UB array is $\sim\SI{2}{\percent}$. The UUB, with its higher resolution, is more susceptible to lightning-induced triggering. The drop in the SD duty cycle for Phase$\,$II reads $5\%$. The algorithm presented in \Cref{sec:rejection} partially fixes this issue.

\section{Improved trigger algorithm}
\label{sec:cleaning}

A new trigger algorithm has been in operation in the UUB electronics since August 2024, which aims to diminish the sensitivity of the T2 algorithms to electronic noise. This is achieved by first filtering symmetric oscillations in the trace and then providing the established triggers with the conditioned version of the detector information.

To detect oscillations, the trace conditioner employs a sliding window algorithm that continuously scans the last 15 samples ($\SI{125}{\nano\second}$) of the trace data. For each sample $b_i$, the algorithm determines the minimum $\wedge$ and the maximum $\vee$ for the three right- and leftmost bins $b_{i\pm7}$, $b_{i\pm6}$, $b_{i\pm5}$. It checks for a sinusoidal pattern in the samples $b_i$, $\left\{\wedge,\,\vee\right\}^\mathrm{right}_\mathrm{left}$ around the baseline $B$ (Ref. \cite{sdcalibration}), that is to say, the conditioner tests whether Eqs. (\ref{eq:trace-cleaning-1}) or (\ref{eq:trace-cleaning-2}) are satisfied.
\begin{alignat}{3}
    \label{eq:trace-cleaning-1}
    \wedge_\mathrm{right} + \SI{4}{ADC} < \, &B \,< b_i, \qquad \wedge_\mathrm{left} + \SI{4}{ADC} <\, &B \,< b_i. 
    \qquad\qquad &\mathrm{Case:}\,\,
    \includegraphics[height=5mm, trim=0 0 0 -1cm]{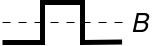}
    \\
    \label{eq:trace-cleaning-2}
    \vee_\mathrm{right} - \SI{4}{ADC} >\, &B \,> b_i, \qquad \vee_\mathrm{left} - \SI{4}{ADC} >\, &B \,> b_i.
    \qquad\qquad &\mathrm{Case:}\,\,
    \includegraphics[height=5mm, trim=0 0 0 -1cm]{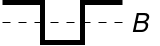}
\end{alignat}

In the Eqs. (\ref{eq:trace-cleaning-1}-\ref{eq:trace-cleaning-2}), a dead band of $2\sigma_B = \SI{4}{ADC}$ counts is employed. Oscillations below this threshold are not considered spurious and are not altered. If a sinusoidal pattern is found, the conditioned trace is calculated according to Eqs. (\ref{eq:conditioned-trace-1}-\ref{eq:conditioned-trace-2}). If no pattern is found, it is $b^\mathrm{cond.}_i = b_i$.

\begin{alignat}{2}
    \label{eq:conditioned-trace-1}
    b^\mathrm{cond.}_i &= b_i + \left(\max(\wedge_\mathrm{left}, \wedge_\mathrm{right}) - B \right) + \SI{0.5}{ADC}
    \qquad\qquad\qquad\qquad &\mathrm{Case:}\,\,
    \includegraphics[height=5mm, trim=0 0 0 -1cm]{fig/up_pulse.png}
    \\
    \label{eq:conditioned-trace-2}
    b^\mathrm{cond.}_i &= b_i + \left(\min(\vee_\mathrm{left}, \vee_\mathrm{right}) - B \right) - \SI{0.5}{ADC}
    \qquad\qquad\qquad\qquad &\mathrm{Case:}\,\,
    \includegraphics[height=5mm, trim=0 0 0 -1cm]{fig/down_pulse.png}
\end{alignat}

\section{Rejection of stale T3s}
\label{sec:rejection}

\begin{figure}
  \centering
  \subfloat[]{\includegraphics[height=0.637\textwidth]{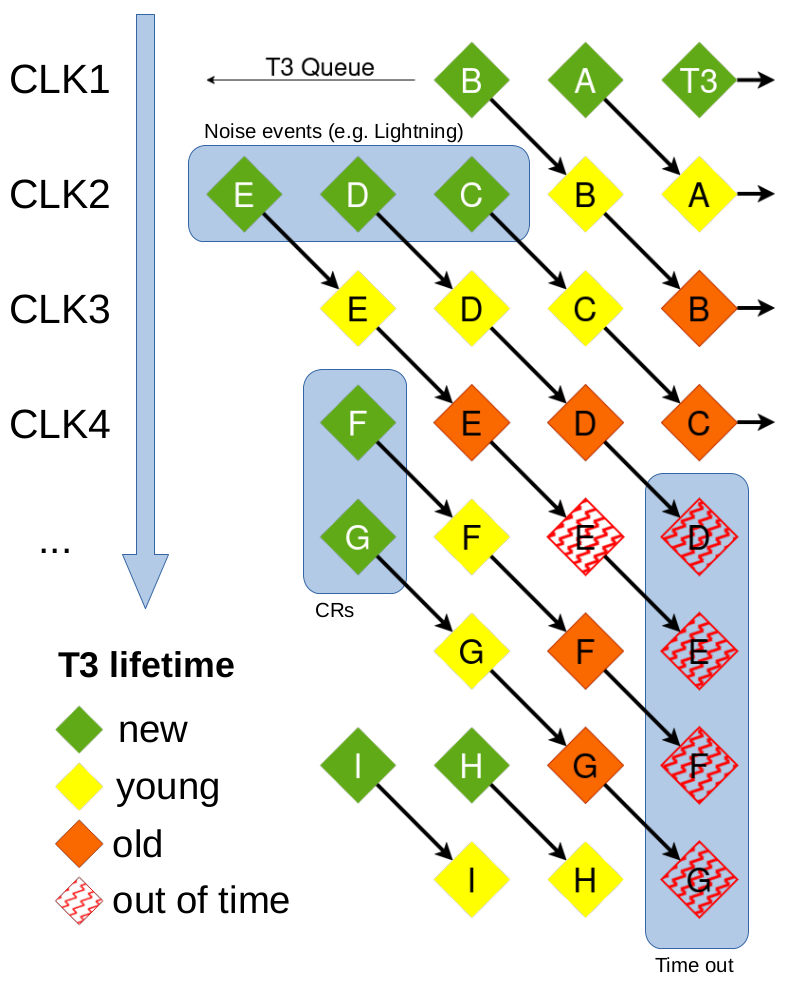}
  \label{fig:t3-queue-old}
  }\hfill
  \subfloat[]{\includegraphics[height=0.663\textwidth]{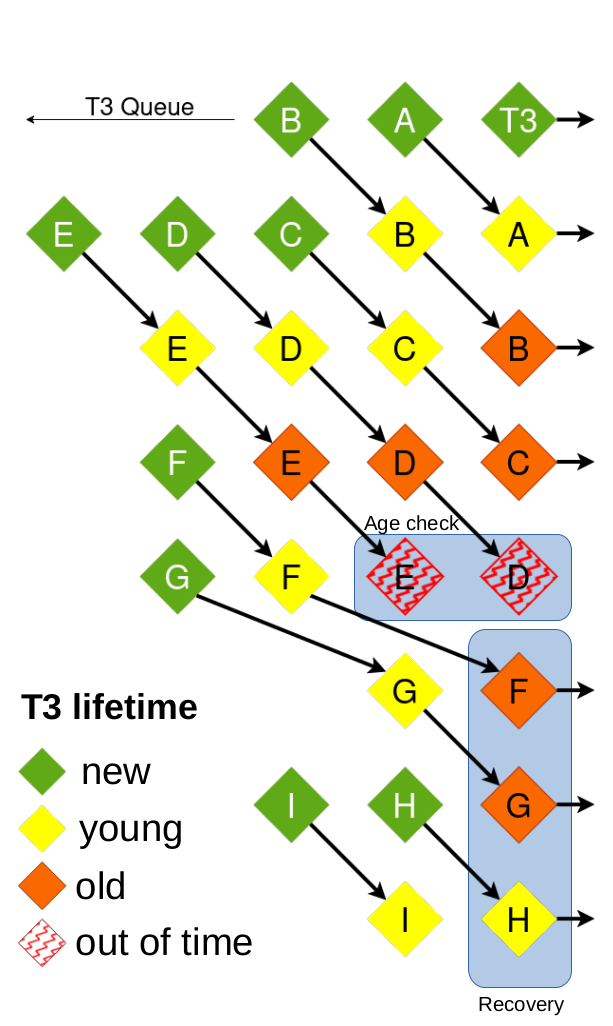}
  \label{fig:t3-queue-new}
  }
  \caption{\subref{fig:t3-queue-old} A large backlog of T3s (caused, e.g., by lightning) can cripple the CDAS performance for much longer than the actual bad weather period. \subref{fig:t3-queue-new} A simple age check rejects T3 requests that are out of time. The CDAS recovers much faster as a consequence.}
  \label{fig:queues}
\end{figure}

An algorithmically simple method for mitigating the effect of stale T3s on the T3 rate and purity has been in operation at the CDAS since January 2025. Before sending a T3 request to stations, a maximum age check is applied at the CDAS level. If the T2 triggers forming the T3 coincidence are older than the cutoff time, the CDAS will reject the request and discard it instead of sending it to the SD array. The schematic operation and its advantages over the naive iteration of the T3 queue are shown in \Cref{fig:queues}. The projected improvement in the duty cycle of the SD array is up to 5\%.

\section{Conclusion}

The Central Data Acquisition System (CDAS) is the main computing facility of the Pierre Auger Observatory. Most importantly, it is responsible for recognizing extensive air shower events from spatio-temporal coincidences of T2  triggers and communicating with candidate stations. The CDAS performed this task reliably during Phase$\,$I of data collection, as shown by several performance parameters that are monitored throughout the operation. These performance parameters remain largely unchanged after the AugerPrime detector upgrade. 

With the higher resolution of detector data, several sources of noise have impacted the event detection quality in non-trivial ways. The effects leading to losses in T3 rates and purities have been studied in depth and solutions have been deployed that mitigate their impact on data quality. The CDAS, in its current state, represents a well-tested system that is prepared for the challenges that come with the AugerPrime detector upgrade and is expected to improve its Phase$\,$I performance.

\newpage
\section*{The Pierre Auger Collaboration}

{\footnotesize\setlength{\baselineskip}{10pt}
\noindent
\begin{wrapfigure}[11]{l}{0.12\linewidth}
\vspace{-4pt}
\includegraphics[width=0.98\linewidth]{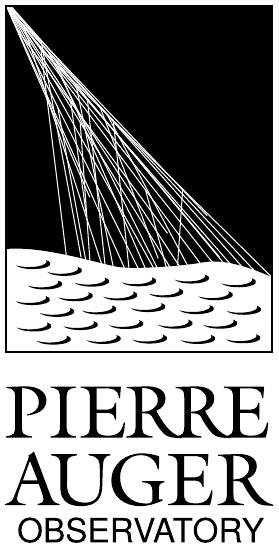}
\end{wrapfigure}
\begin{sloppypar}\noindent
A.~Abdul Halim$^{13}$,
P.~Abreu$^{70}$,
M.~Aglietta$^{53,51}$,
I.~Allekotte$^{1}$,
K.~Almeida Cheminant$^{78,77}$,
A.~Almela$^{7,12}$,
R.~Aloisio$^{44,45}$,
J.~Alvarez-Mu\~niz$^{76}$,
A.~Ambrosone$^{44}$,
J.~Ammerman Yebra$^{76}$
% ...

\end{sloppypar}
\begin{center}
\end{center}

\vspace{1ex}
\begin{description}[labelsep=0.2em,align=right,labelwidth=0.7em,labelindent=0em,leftmargin=2em,noitemsep,before={\renewcommand\makelabel[1]{##1 }}]
\item[$^{1}$] Centro At\'omico Bariloche and Instituto Balseiro (CNEA-UNCuyo-CONICET), San Carlos de Bariloche, Argentina
\item[$^{2}$] Departamento de F\'\i{}sica and Departamento de Ciencias de la Atm\'osfera y los Oc\'eanos, FCEyN, Universidad de Buenos Aires and CONICET, Buenos Aires, Argentina
\item[$^{3}$] IFLP, Universidad Nacional de La Plata and CONICET, La Plata, Argentina
\item[$^{4}$] Instituto de Astronom\'\i{}a y F\'\i{}sica del Espacio (IAFE, CONICET-UBA), Buenos Aires, Argentina
\item[$^{5}$] Instituto de F\'\i{}sica de Rosario (IFIR) -- CONICET/U.N.R.\ and Facultad de Ciencias Bioqu\'\i{}micas y Farmac\'euticas U.N.R., Rosario, Argentina
\item[$^{6}$] Instituto de Tecnolog\'\i{}as en Detecci\'on y Astropart\'\i{}culas (CNEA, CONICET, UNSAM), and Universidad Tecnol\'ogica Nacional -- Facultad Regional Mendoza (CONICET/CNEA), Mendoza, Argentina
\item[$^{7}$] Instituto de Tecnolog\'\i{}as en Detecci\'on y Astropart\'\i{}culas (CNEA, CONICET, UNSAM), Buenos Aires, Argentina
\item[$^{8}$] International Center of Advanced Studies and Instituto de Ciencias F\'\i{}sicas, ECyT-UNSAM and CONICET, Campus Miguelete -- San Mart\'\i{}n, Buenos Aires, Argentina
\item[$^{9}$] Laboratorio Atm\'osfera -- Departamento de Investigaciones en L\'aseres y sus Aplicaciones -- UNIDEF (CITEDEF-CONICET), Argentina
\item[$^{10}$] Observatorio Pierre Auger, Malarg\"ue, Argentina
\item[$^{11}$] Observatorio Pierre Auger and Comisi\'on Nacional de Energ\'\i{}a At\'omica, Malarg\"ue, Argentina
\item[$^{12}$] Universidad Tecnol\'ogica Nacional -- Facultad Regional Buenos Aires, Buenos Aires, Argentina
\item[$^{13}$] University of Adelaide, Adelaide, S.A., Australia
\item[$^{14}$] Universit\'e Libre de Bruxelles (ULB), Brussels, Belgium
\item[$^{15}$] Vrije Universiteit Brussels, Brussels, Belgium
\item[$^{16}$] Centro Brasileiro de Pesquisas Fisicas, Rio de Janeiro, RJ, Brazil
\item[$^{17}$] Centro Federal de Educa\c{c}\~ao Tecnol\'ogica Celso Suckow da Fonseca, Petropolis, Brazil
\item[$^{18}$] Instituto Federal de Educa\c{c}\~ao, Ci\^encia e Tecnologia do Rio de Janeiro (IFRJ), Brazil
\item[$^{19}$] Universidade de S\~ao Paulo, Escola de Engenharia de Lorena, Lorena, SP, Brazil
\item[$^{20}$] Universidade de S\~ao Paulo, Instituto de F\'\i{}sica de S\~ao Carlos, S\~ao Carlos, SP, Brazil
\item[$^{21}$] Universidade de S\~ao Paulo, Instituto de F\'\i{}sica, S\~ao Paulo, SP, Brazil
\item[$^{22}$] Universidade Estadual de Campinas (UNICAMP), IFGW, Campinas, SP, Brazil
\item[$^{23}$] Universidade Estadual de Feira de Santana, Feira de Santana, Brazil
\item[$^{24}$] Universidade Federal de Campina Grande, Centro de Ciencias e Tecnologia, Campina Grande, Brazil
\item[$^{25}$] Universidade Federal do ABC, Santo Andr\'e, SP, Brazil
\item[$^{26}$] Universidade Federal do Paran\'a, Setor Palotina, Palotina, Brazil
\item[$^{27}$] Universidade Federal do Rio de Janeiro, Instituto de F\'\i{}sica, Rio de Janeiro, RJ, Brazil
\item[$^{28}$] Universidad de Medell\'\i{}n, Medell\'\i{}n, Colombia
\item[$^{29}$] Universidad Industrial de Santander, Bucaramanga, Colombia
\item[$^{30}$] Charles University, Faculty of Mathematics and Physics, Institute of Particle and Nuclear Physics, Prague, Czech Republic
\item[$^{31}$] Institute of Physics of the Czech Academy of Sciences, Prague, Czech Republic
\item[$^{32}$] Palacky University, Olomouc, Czech Republic
\item[$^{33}$] CNRS/IN2P3, IJCLab, Universit\'e Paris-Saclay, Orsay, France
\item[$^{34}$] Laboratoire de Physique Nucl\'eaire et de Hautes Energies (LPNHE), Sorbonne Universit\'e, Universit\'e de Paris, CNRS-IN2P3, Paris, France
\item[$^{35}$] Univ.\ Grenoble Alpes, CNRS, Grenoble Institute of Engineering Univ.\ Grenoble Alpes, LPSC-IN2P3, 38000 Grenoble, France
\item[$^{36}$] Universit\'e Paris-Saclay, CNRS/IN2P3, IJCLab, Orsay, France
\item[$^{37}$] Bergische Universit\"at Wuppertal, Department of Physics, Wuppertal, Germany
\item[$^{38}$] Karlsruhe Institute of Technology (KIT), Institute for Experimental Particle Physics, Karlsruhe, Germany
\item[$^{39}$] Karlsruhe Institute of Technology (KIT), Institut f\"ur Prozessdatenverarbeitung und Elektronik, Karlsruhe, Germany
\item[$^{40}$] Karlsruhe Institute of Technology (KIT), Institute for Astroparticle Physics, Karlsruhe, Germany
\item[$^{41}$] RWTH Aachen University, III.\ Physikalisches Institut A, Aachen, Germany
\item[$^{42}$] Universit\"at Hamburg, II.\ Institut f\"ur Theoretische Physik, Hamburg, Germany
\item[$^{43}$] Universit\"at Siegen, Department Physik -- Experimentelle Teilchenphysik, Siegen, Germany
\item[$^{44}$] Gran Sasso Science Institute, L'Aquila, Italy
\item[$^{45}$] INFN Laboratori Nazionali del Gran Sasso, Assergi (L'Aquila), Italy
\item[$^{46}$] INFN, Sezione di Catania, Catania, Italy
\item[$^{47}$] INFN, Sezione di Lecce, Lecce, Italy
\item[$^{48}$] INFN, Sezione di Milano, Milano, Italy
\item[$^{49}$] INFN, Sezione di Napoli, Napoli, Italy
\item[$^{50}$] INFN, Sezione di Roma ``Tor Vergata'', Roma, Italy
\item[$^{51}$] INFN, Sezione di Torino, Torino, Italy
\item[$^{52}$] Istituto di Astrofisica Spaziale e Fisica Cosmica di Palermo (INAF), Palermo, Italy
\item[$^{53}$] Osservatorio Astrofisico di Torino (INAF), Torino, Italy
\item[$^{54}$] Politecnico di Milano, Dipartimento di Scienze e Tecnologie Aerospaziali , Milano, Italy
\item[$^{55}$] Universit\`a del Salento, Dipartimento di Matematica e Fisica ``E.\ De Giorgi'', Lecce, Italy
\item[$^{56}$] Universit\`a dell'Aquila, Dipartimento di Scienze Fisiche e Chimiche, L'Aquila, Italy
\item[$^{57}$] Universit\`a di Catania, Dipartimento di Fisica e Astronomia ``Ettore Majorana``, Catania, Italy
\item[$^{58}$] Universit\`a di Milano, Dipartimento di Fisica, Milano, Italy
\item[$^{59}$] Universit\`a di Napoli ``Federico II'', Dipartimento di Fisica ``Ettore Pancini'', Napoli, Italy
\item[$^{60}$] Universit\`a di Palermo, Dipartimento di Fisica e Chimica ''E.\ Segr\`e'', Palermo, Italy
\item[$^{61}$] Universit\`a di Roma ``Tor Vergata'', Dipartimento di Fisica, Roma, Italy
\item[$^{62}$] Universit\`a Torino, Dipartimento di Fisica, Torino, Italy
\item[$^{63}$] Benem\'erita Universidad Aut\'onoma de Puebla, Puebla, M\'exico
\item[$^{64}$] Unidad Profesional Interdisciplinaria en Ingenier\'\i{}a y Tecnolog\'\i{}as Avanzadas del Instituto Polit\'ecnico Nacional (UPIITA-IPN), M\'exico, D.F., M\'exico
\item[$^{65}$] Universidad Aut\'onoma de Chiapas, Tuxtla Guti\'errez, Chiapas, M\'exico
\item[$^{66}$] Universidad Michoacana de San Nicol\'as de Hidalgo, Morelia, Michoac\'an, M\'exico
\item[$^{67}$] Universidad Nacional Aut\'onoma de M\'exico, M\'exico, D.F., M\'exico
\item[$^{68}$] Institute of Nuclear Physics PAN, Krakow, Poland
\item[$^{69}$] University of \L{}\'od\'z, Faculty of High-Energy Astrophysics,\L{}\'od\'z, Poland
\item[$^{70}$] Laborat\'orio de Instrumenta\c{c}\~ao e F\'\i{}sica Experimental de Part\'\i{}culas -- LIP and Instituto Superior T\'ecnico -- IST, Universidade de Lisboa -- UL, Lisboa, Portugal
\item[$^{71}$] ``Horia Hulubei'' National Institute for Physics and Nuclear Engineering, Bucharest-Magurele, Romania
\item[$^{72}$] Institute of Space Science, Bucharest-Magurele, Romania
\item[$^{73}$] Center for Astrophysics and Cosmology (CAC), University of Nova Gorica, Nova Gorica, Slovenia
\item[$^{74}$] Experimental Particle Physics Department, J.\ Stefan Institute, Ljubljana, Slovenia
\item[$^{75}$] Universidad de Granada and C.A.F.P.E., Granada, Spain
\item[$^{76}$] Instituto Galego de F\'\i{}sica de Altas Enerx\'\i{}as (IGFAE), Universidade de Santiago de Compostela, Santiago de Compostela, Spain
\item[$^{77}$] IMAPP, Radboud University Nijmegen, Nijmegen, The Netherlands
\item[$^{78}$] Nationaal Instituut voor Kernfysica en Hoge Energie Fysica (NIKHEF), Science Park, Amsterdam, The Netherlands
\item[$^{79}$] Stichting Astronomisch Onderzoek in Nederland (ASTRON), Dwingeloo, The Netherlands
\item[$^{80}$] Universiteit van Amsterdam, Faculty of Science, Amsterdam, The Netherlands
\item[$^{81}$] Case Western Reserve University, Cleveland, OH, USA
\item[$^{82}$] Colorado School of Mines, Golden, CO, USA
\item[$^{83}$] Department of Physics and Astronomy, Lehman College, City University of New York, Bronx, NY, USA
\item[$^{84}$] Michigan Technological University, Houghton, MI, USA
\item[$^{85}$] New York University, New York, NY, USA
\item[$^{86}$] University of Chicago, Enrico Fermi Institute, Chicago, IL, USA
\item[$^{87}$] University of Delaware, Department of Physics and Astronomy, Bartol Research Institute, Newark, DE, USA
\item[] -----
\item[$^{a}$] Max-Planck-Institut f\"ur Radioastronomie, Bonn, Germany
\item[$^{b}$] also at Kapteyn Institute, University of Groningen, Groningen, The Netherlands
\item[$^{c}$] School of Physics and Astronomy, University of Leeds, Leeds, United Kingdom
\item[$^{d}$] Fermi National Accelerator Laboratory, Fermilab, Batavia, IL, USA
\item[$^{e}$] Pennsylvania State University, University Park, PA, USA
\item[$^{f}$] Colorado State University, Fort Collins, CO, USA
\item[$^{g}$] Louisiana State University, Baton Rouge, LA, USA
\item[$^{h}$] now at Graduate School of Science, Osaka Metropolitan University, Osaka, Japan
\item[$^{i}$] Institut universitaire de France (IUF), France
\item[$^{j}$] now at Technische Universit\"at Dortmund and Ruhr-Universit\"at Bochum, Dortmund and Bochum, Germany
\end{description}

\newpage
\section*{Acknowledgments}

\begin{sloppypar}
The successful installation, commissioning, and operation of the Pierre
Auger Observatory would not have been possible without the strong
commitment and effort from the technical and administrative staff in
Malarg\"ue. We are very grateful to the following agencies and
organizations for financial support:
\end{sloppypar}

\begin{sloppypar}
Argentina -- Comisi\'on Nacional de Energ\'\i{}a At\'omica; Agencia Nacional de
Promoci\'on Cient\'\i{}fica y Tecnol\'ogica (ANPCyT); Consejo Nacional de
Investigaciones Cient\'\i{}ficas y T\'ecnicas (CONICET); Gobierno de la
Provincia de Mendoza; Municipalidad de Malarg\"ue; NDM Holdings and Valle
Las Le\~nas; in gratitude for their continuing cooperation over land
access; Australia -- the Australian Research Council; Belgium -- Fonds
de la Recherche Scientifique (FNRS); Research Foundation Flanders (FWO),
Marie Curie Action of the European Union Grant No.~101107047; Brazil --
Conselho Nacional de Desenvolvimento Cient\'\i{}fico e Tecnol\'ogico (CNPq);
Financiadora de Estudos e Projetos (FINEP); Funda\c{c}\~ao de Amparo \`a
Pesquisa do Estado de Rio de Janeiro (FAPERJ); S\~ao Paulo Research
Foundation (FAPESP) Grants No.~2019/10151-2, No.~2010/07359-6 and
No.~1999/05404-3; Minist\'erio da Ci\^encia, Tecnologia, Inova\c{c}\~oes e
Comunica\c{c}\~oes (MCTIC); Czech Republic -- GACR 24-13049S, CAS LQ100102401,
MEYS LM2023032, CZ.02.1.01/0.0/0.0/16{\textunderscore}013/0001402,
CZ.02.1.01/0.0/0.0/18{\textunderscore}046/0016010 and
CZ.02.1.01/0.0/0.0/17{\textunderscore}049/0008422 and CZ.02.01.01/00/22{\textunderscore}008/0004632;
France -- Centre de Calcul IN2P3/CNRS; Centre National de la Recherche
Scientifique (CNRS); Conseil R\'egional Ile-de-France; D\'epartement
Physique Nucl\'eaire et Corpusculaire (PNC-IN2P3/CNRS); D\'epartement
Sciences de l'Univers (SDU-INSU/CNRS); Institut Lagrange de Paris (ILP)
Grant No.~LABEX ANR-10-LABX-63 within the Investissements d'Avenir
Programme Grant No.~ANR-11-IDEX-0004-02; Germany -- Bundesministerium
f\"ur Bildung und Forschung (BMBF); Deutsche Forschungsgemeinschaft (DFG);
Finanzministerium Baden-W\"urttemberg; Helmholtz Alliance for
Astroparticle Physics (HAP); Helmholtz-Gemeinschaft Deutscher
Forschungszentren (HGF); Ministerium f\"ur Kultur und Wissenschaft des
Landes Nordrhein-Westfalen; Ministerium f\"ur Wissenschaft, Forschung und
Kunst des Landes Baden-W\"urttemberg; Italy -- Istituto Nazionale di
Fisica Nucleare (INFN); Istituto Nazionale di Astrofisica (INAF);
Ministero dell'Universit\`a e della Ricerca (MUR); CETEMPS Center of
Excellence; Ministero degli Affari Esteri (MAE), ICSC Centro Nazionale
di Ricerca in High Performance Computing, Big Data and Quantum
Computing, funded by European Union NextGenerationEU, reference code
CN{\textunderscore}00000013; M\'exico -- Consejo Nacional de Ciencia y Tecnolog\'\i{}a
(CONACYT) No.~167733; Universidad Nacional Aut\'onoma de M\'exico (UNAM);
PAPIIT DGAPA-UNAM; The Netherlands -- Ministry of Education, Culture and
Science; Netherlands Organisation for Scientific Research (NWO); Dutch
national e-infrastructure with the support of SURF Cooperative; Poland
-- Ministry of Education and Science, grants No.~DIR/WK/2018/11 and
2022/WK/12; National Science Centre, grants No.~2016/22/M/ST9/00198,
2016/23/B/ST9/01635, 2020/39/B/ST9/01398, and 2022/45/B/ST9/02163;
Portugal -- Portuguese national funds and FEDER funds within Programa
Operacional Factores de Competitividade through Funda\c{c}\~ao para a Ci\^encia
e a Tecnologia (COMPETE); Romania -- Ministry of Research, Innovation
and Digitization, CNCS-UEFISCDI, contract no.~30N/2023 under Romanian
National Core Program LAPLAS VII, grant no.~PN 23 21 01 02 and project
number PN-III-P1-1.1-TE-2021-0924/TE57/2022, within PNCDI III; Slovenia
-- Slovenian Research Agency, grants P1-0031, P1-0385, I0-0033, N1-0111;
Spain -- Ministerio de Ciencia e Innovaci\'on/Agencia Estatal de
Investigaci\'on (PID2019-105544GB-I00, PID2022-140510NB-I00 and
RYC2019-027017-I), Xunta de Galicia (CIGUS Network of Research Centers,
Consolidaci\'on 2021 GRC GI-2033, ED431C-2021/22 and ED431F-2022/15),
Junta de Andaluc\'\i{}a (SOMM17/6104/UGR and P18-FR-4314), and the European
Union (Marie Sklodowska-Curie 101065027 and ERDF); USA -- Department of
Energy, Contracts No.~DE-AC02-07CH11359, No.~DE-FR02-04ER41300,
No.~DE-FG02-99ER41107 and No.~DE-SC0011689; National Science Foundation,
Grant No.~0450696, and NSF-2013199; The Grainger Foundation; Marie
Curie-IRSES/EPLANET; European Particle Physics Latin American Network;
and UNESCO.
\end{sloppypar}

}


\newpage
\begin{thebibliography}{99}

\setlength{\parskip}{1pt}
\setlength{\itemsep}{0pt plus 0.3ex}

\bibitem{satoProceeding}
    R. Sato [Pierre Auger Collaboration],
    \href{https://doi.org/10.22323/1.444.0373}{%
    PoS \textbf{ICRC2023} (2023) 373.
    }

\bibitem{sdtriggerpaper}
    J. Abraham \emph{et al.},
    \href{https://doi.org/10.1016/j.nima.2009.11.018}{%
    Nucl. Instrum. Methods A, \textbf{613} (2010) 029-039.
    }

\bibitem{augerprime}
    A. Castellina [Pierre Auger Collaboration], 
    \href{https://doi.org/10.1051/epjconf/201921006002}{%
    EPJ Web Conf. \textbf{210} (2019) 06002.
    }

\bibitem{carlamonitoring}
    C. Bonifazi [Pierre Auger Collaboration],
    \href{https://doi.org/10.48550/arXiv.1307.5059}{%
    JINST \textbf{ICRC2013} (2013) 019-022.
    }

\bibitem{sdcalibration}
    X. Bertou \emph{et al.},
    \href{https://doi.org/10.1016/j.nima.2006.07.066}{%
    Nucl. Instrum. Methods A, \textbf{568} (2006) 839-846.
    }

\end{thebibliography}
\end{document}